# Protons in lattice confinement: Static pressure on the Y-substituted, hydrated $BaZrO_3$ ceramic proton conductor decreases proton mobility

Qianli Chen[1,2], Artur Braun[1*], Alejandro Ovalle[1], Cristian-Daniel Savaniu[3], Thomas Graule[1, 4], Nikolai Bagdassarov[5]

[1]*Laboratory for High Performance Ceramics*

*EMPA – Swiss Federal Laboratories for Materials Testing and Research*

*CH-8600 Dübendorf, Switzerland*

[2] *Department of Physics, ETH Zürich, Swiss Federal Institute of Technology*

*CH-8057 Zürich, Switzerland*

[3]*University of St. Andrews, School of Chemistry*

*St Andrews, Fife, KY16 8DA, Scotland, United Kingdom*

[4]*Technische Universität Bergakademie Freiberg, D-09596 Freiberg, Germany*

[5] *Institut for Geosciences, J. W. Goethe Universität Frankfurt am Main*

*D-60323 Frankfurt/Main, Germany*

[*] Corresponding author: Phone +41 44 823 4850, Fax: +41 44 823 4150,

email artur.braun@alumni.ethz.ch

ABSTRACT

Yttrium substituted $BaZrO_3$, with nominal composition BaZr0.9Y0.1O3, a ceramic proton conductor, was subject to impedance spectroscopy for temperatures 300 K < T < 715 K at mechanical pressures 1 GPa < p < 2 GPa. The activation energies $E_a$ of bulk and grain boundary conductivity from two perovskites synthesized by solid-state reaction and sol-gel method were determined under high pressures. At high temperature, the bulk activation energy increases with pressure by 5% for sol-gel derived sample and by 40% for solid-state derived sample. For the sample prepared by solid-state reaction, there is a large gap of 0.17 eV between the activation energy at 1.0 GPa and > 1.2 GPa. The grain boundary activation energy is around a factor two times as that of the bulk, and it reaches a maximum at 1.25 – 1.5 GPa, and then decrease as the pressure increases, indicating higher proton mobility in the grain boundaries at higher pressure. Since this effect is not reversible, it is suggested that the grain boundary resistance decreases as a result of pressure induced sintering. The steady increase of the bulk resistivtiy upon pressurising suggests that the proton mobility depends on the space available in the lattice. In return, an expanded lattice with a/a0 > 1 should thus have a lower activaqtion energy, suggesting that thin films expansive tensile strain could have a larger proton conductivity with desirable properties for applications.

Perovskite-type ceramics with proton conducting properties have the potential for a wide variety of applications. For example, they can be employed as solid electrolyte materials in the design of electrochemical devices such as fuel cells. Comparing to conventional solid electrolytes which are based on oxygen ion conductivity, materials with pronounced proton conductivity are attractive for lowering the operating temperature. $BaZr_{0.9}Y_{0.1}O_{2.95}$ (BZY10), in which oxygen vacancies are created by Y-doping, is an attractive candidate material [references]. While the fundamental physical and chemical processes of the proton conductivity are not all well known, the current opinion is that in wet atmosphere, the vacancy can be filled by oxygen which dissociates from adsorbed water molecule[1], and the introduced protons are bound to the interstitial lattice oxygen[2]. Protons reorient in high frequency locally around a lattice oxygen ion with activation energy in the order of few meV at moderate temperature[3]; whereas at elevated temperature, thermal activation may drive the proton to another oxygen site by overcoming the energy barrier in the range of 0.1 to 1.0 eV[4], and thus constituting charge mobility. For improving device applications it is of high interest to understand the relationship of structure parameter and proton transport properties. The potential relations between the activation energy and structure parameters were pointed out as early as 1988[5]. Our recent findings modifying “chemical pressure” - achieved by different synthesis routes - suggest that $E_a$ increases with decreasing lattice spacing[6,7]. It was confirmed by Ashok[8] et al. that an expansion of unit cell caused by vary the concentration of the A-site element eases the carrier mobility. Interestingly, Wu et al[9,10] reported that heavily Y-doped barium cerate exhibits significantly lower conductivity than those of lightly doped analogues, contradicting those findings. While change of the A-site cation size has a direct effect on the lattice volume and also on the proton conductivity, we cannot generally rule out that the different chemical nature of the chosen substituents has an additional specific influence on the proton conductivity. To investigate intuitively the change in material lattice, it is technically more challenging to expand the lattice volume. For the immediate and ultimate proof of the aforementioned hypothesis, the direct method is to apply mechanical pressure- and temperature-dependent analytical techniques to monitor the proton transport. We pay attention to the possibility that high pressure may affect the grain boundary structure, which modifies the proton conductivity. In this work, 10% Yttrium substituted $BaZrO_3$ was prepared and the structure determined by X-ray diffraction. After hydration, Their electrical conductivities at high pressure and high temperature were studied to have a better understanding for the nature of proton transport in those materials.

Two batches of $BaZr_{0.9}Y_{0.1}O_{2.95}$ were synthesized, correspondingly by solid-state reaction (SS) [11,12] and sol-gel method (SG). Sol-gel material was prepared starting from barium acetate (99%), zirconium (IV) 2,4-pentanedionate (99%) and yttrium (III) 2-ethylhexanoate (99.8%)-all from Alfa Aesar- acetic acid being used as chelating agent and solvent. The reactants were dissolved in appropriate ratios in acetic acid at 80°C and the resulting clear sol was stirred for another 3 hours, followed by slow evaporation on a hot plate, at 150°C. The resultant dark brown syrup remained precipitate-free, leading to very fine powder by heating to 400°C and then to 1200°C to ensure the formation of the corresponding BZY10 oxide. Pellets of material were pressed and fired at 1500°C for 12 hours for densification.

X-ray diffraction and preliminary Rietveld refinement (see Figure 1) showed that both samples prepared by solid-state and sol-gel routes had cubic structures *Pm3m* ($a_{SS}$ = 4.2111 Å and $a_{SG}$ = 4.2025 Å) after sintering. Close inspection of the diffractogram shows that the solid state derived sample has double peaks and resembles therefore either two cubic similar samples or one tetragonal phase [13]. Scherrer calculation showed the average crystallite size for solid-state derived samples was 60 - 100 nm, and 30 - 40 nm for the sol-gel derived sample.

The hydration process, also referred to as protonation, was described in ref. [2] [2]. Electrical impedance spectra (EIS) were obtained by Solartron 1260 Phase-Gain-Analyser, in the frequency range between 0.01 Hz and 0.5 MHz. A piston-cylinder apparatus[14] was applied on protonated BZY10 powders, at high pressure between 1 GPa and 2 GPa and temperature ranging from 300 K to 680 K. The analysis of impedance data follows our previous works[7,11,12]. Based on recent quasi-elastic neutron scattering, a technique specifically sentitive to protons, it can be stated that the conductivity measured in this temperature range is primarily based on proton conductivity, rather than oxygen conductivity.

The temperature and pressure dependence of the bulk and grain boundary (GB) conductivities for BZY10 prepared by solid-state and sol-gel route were shown respectively in Figure 2 (a) and (b). Particularly the low temperatures (e.g. below 474 K) show that lower pressures provide a higher conductivity than the higher pressures, suggesting that “protons need space”. This effect seems more pronounced for the jump rotational mode (low T) than for the jump diffusion (high T). Figure 3 compares the bulk conductivity at 474 K, 575 K and 671 K for all pressures applied. For the sol-gel sample we find that the grain boundary conductivity is one or two orders of magnitude higher than the bulk conductivity.

The activation energy for bulk and grain boundary conductivities, shown in Figure 4, was determined by Arrhenius plot to the conductivity data shown in Figure 3. At low temperatures $T < 474$ K, smaller activation energies were found; whereas at high temperatures the increase of activation energy indicated that the protons need to overcome an energy barrier to be conductive. The grain boundary activation energies in Figure 4 are always larger than those of the bulk conductivities. For the solid-state derived sample, the activation energy between 400 K and 680 K for bulk conductivity is 0.67 eV for 1 GPa and 0.94 eV for 2 GPa. The activation energy is 0.77 - 0.80 eV for the sol-gel derived sample. The increase in activation energy upon increasing pressure ranges about 5%. There is no intermediate conductivity (e.g. 0.67 eV) as it was noticed by the solid-state derived sample. We cannot entirely rule out that at the beginning of the experiment the solid state derived sample possesses higher water content and higher electrical conductivity than the sol-gel sample. During the first heating cycle the sample may have lost some water and the electrical conductivity became slightly below the conductivity of the sol-gel sample. The subsequent heating and cooling are absolutely reversible. The pressure produces a minor effect on the high temperature conductivity and decrease the conductivity at $T < 170°C$.

Based on data for BaZrO3 by Kurosaki[15] et al., we convert pressure into lattice constant. We assume that the same pressure-dependent lattice parameter expansion applies for BZY10 at elevated temperature, thus it is possible to correlate the conductivity properties with the material lattice constant. Figure 5 shows that for both batches of samples, the lowest activation energy appears with the largest lattice constant realized by the lowest pressure. Applying mechanical pressure has similar effect with tuning the "chemical pressure" (shown as green circles in Figure 5) induced by different synthesis routes and sintering temperatures, as found previously. It was confirmed that the activation energies decrease linearly with the increase of lattice constant. For the sol-gel sample, the extent of lattice constant dependency is, however, not as large as found in the thesis of Empa[7].

The changes in activation energy in the grain boundary, both for solid-state and sol-gel derived samples, are much more significant than those of bulk activation energy, therefore suggesting that grain boundary tailoring is the most efficient way to enhance proton conductivity. A maximum value appeared for the grain boundary activation energies at $p$ = 1.25 GPa or 1.5 GPa. At higher pressure up to 2 GPa, the activation energy decreased to smaller values, indicating an enhancement in electrical conductivity. This is possibly because of the high pressure "sintering effect" on the structure of grain boundary regions. For the sol-gel derived samples, which have smaller crystalline size, this effect is more significant as observed (~32% decrease in $E_a$) comparing to the solid-state derived sample with larger crystallite size (~13% decrease in $E_a$). Under high pressure, electrical conductivity may happen in the vicinity of grain boundaries.

Considering the variation of the activation energy  in the bulk as a function of reduced lattice parameter $a/a_0$, it is suggested that materials under tensile strain, i.e. $a/a_0$ >1, would have an even lower activation energy and thus better proton conductivity. such situation could be realized in epitaxially strained film,s for example.

## Acknowledgment

Funding by E.U. MIRG # CT-2006-042095, Swiss NSF # 200021-124812, Swiss Federal Office of Energy project No. 100411, and EMPA Director's Fund 6th F&E Series.

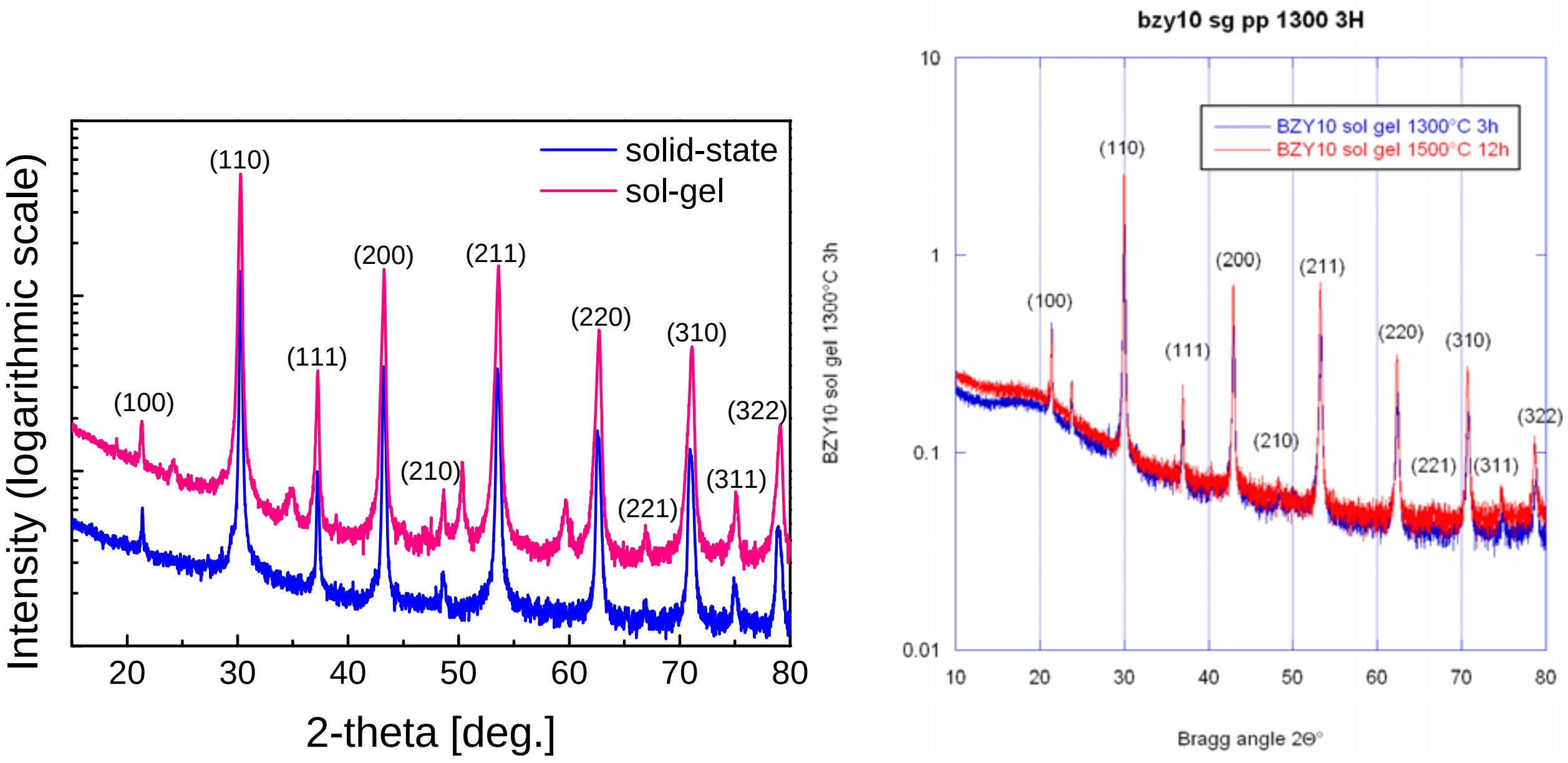

**Figure 1 X-ray powder diffractograms of BZY10 prepared by solid-state reaction (blue) and sol-gel route (red).**

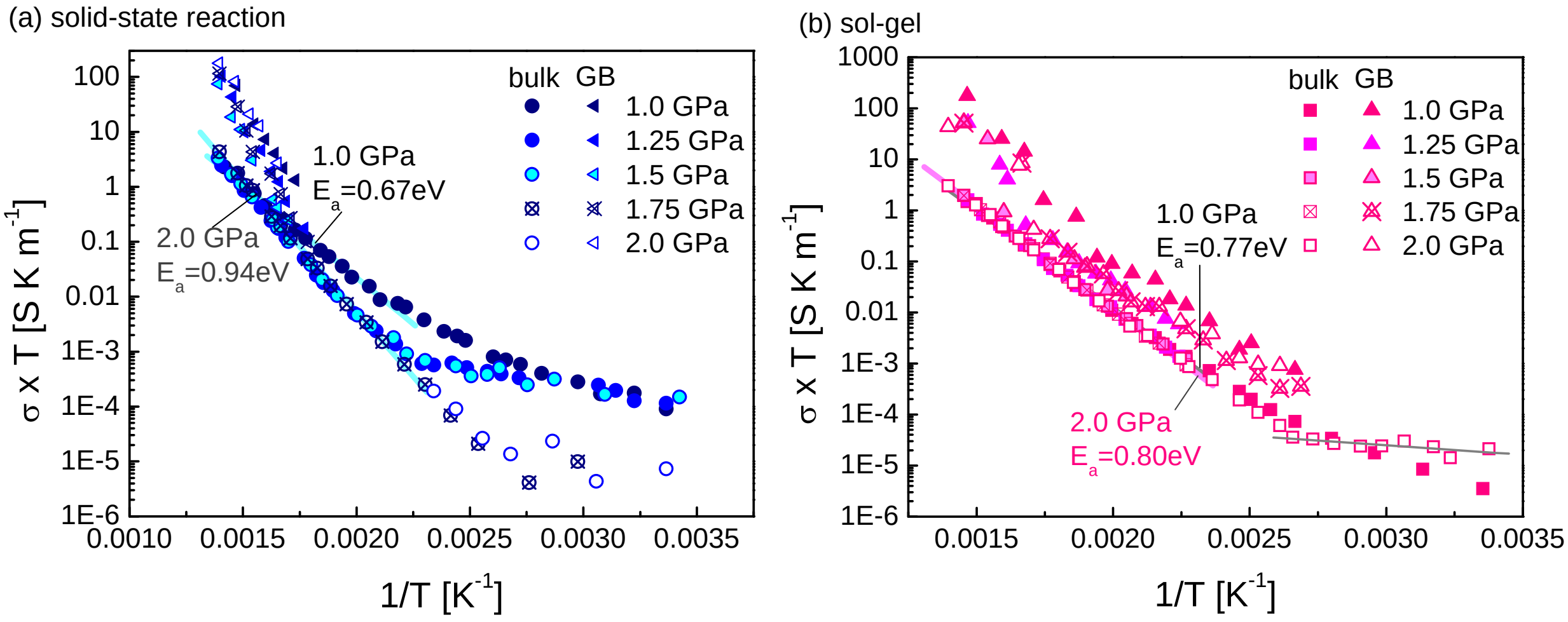

**Figure 2 Conductivity of BZY10 at high pressure and high temperature (a) bulk conductivity (circle) for solid-state derived samples, (b) bulk (square) and grain boundary (triangle) conductivity for sol-gel derived samples.**

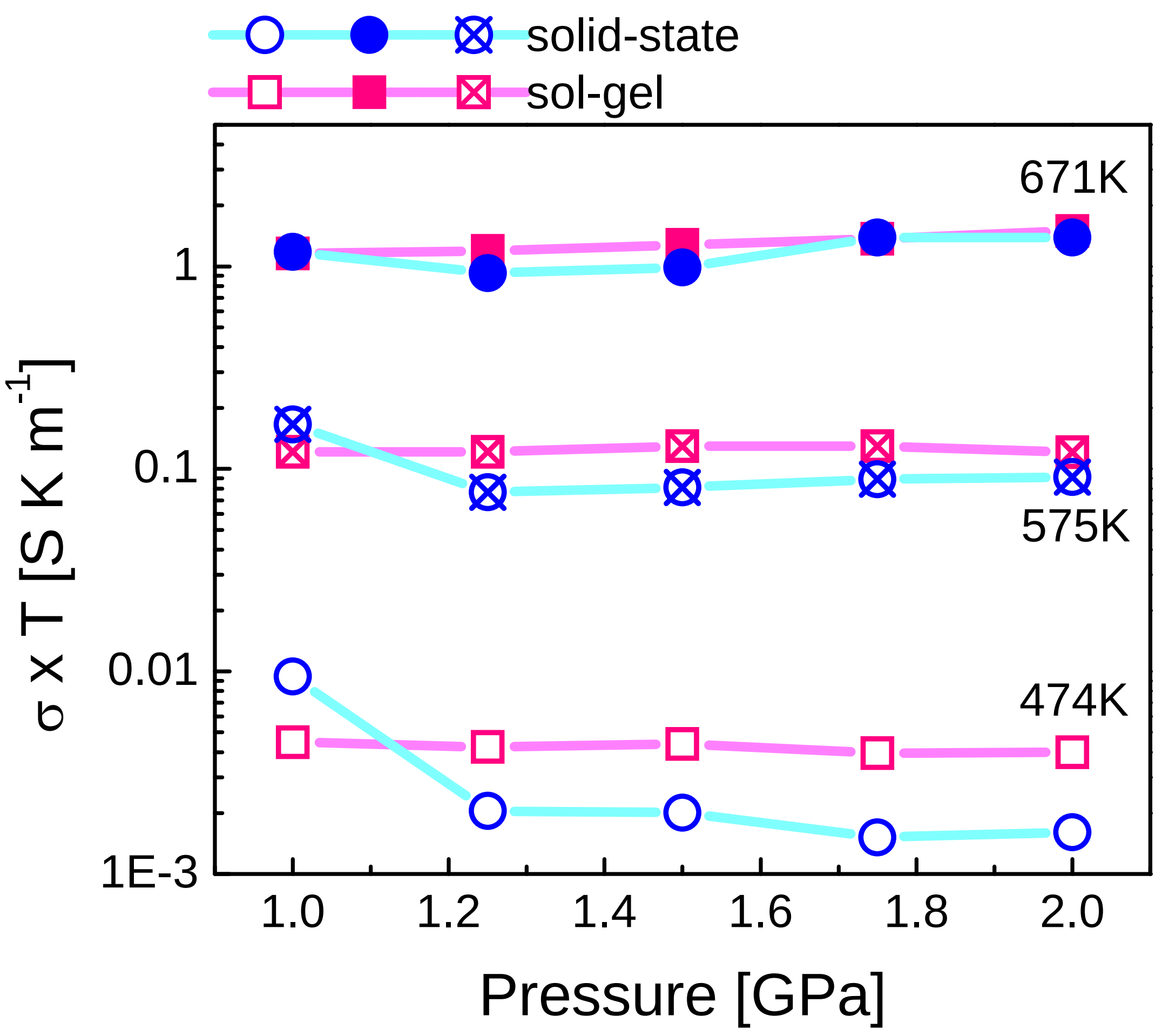

**Figure 3 Bulk conductivity of BZY10 at 474 K, 575 K and 671 K at pressures 1 GPa < p < 2 GPa.**

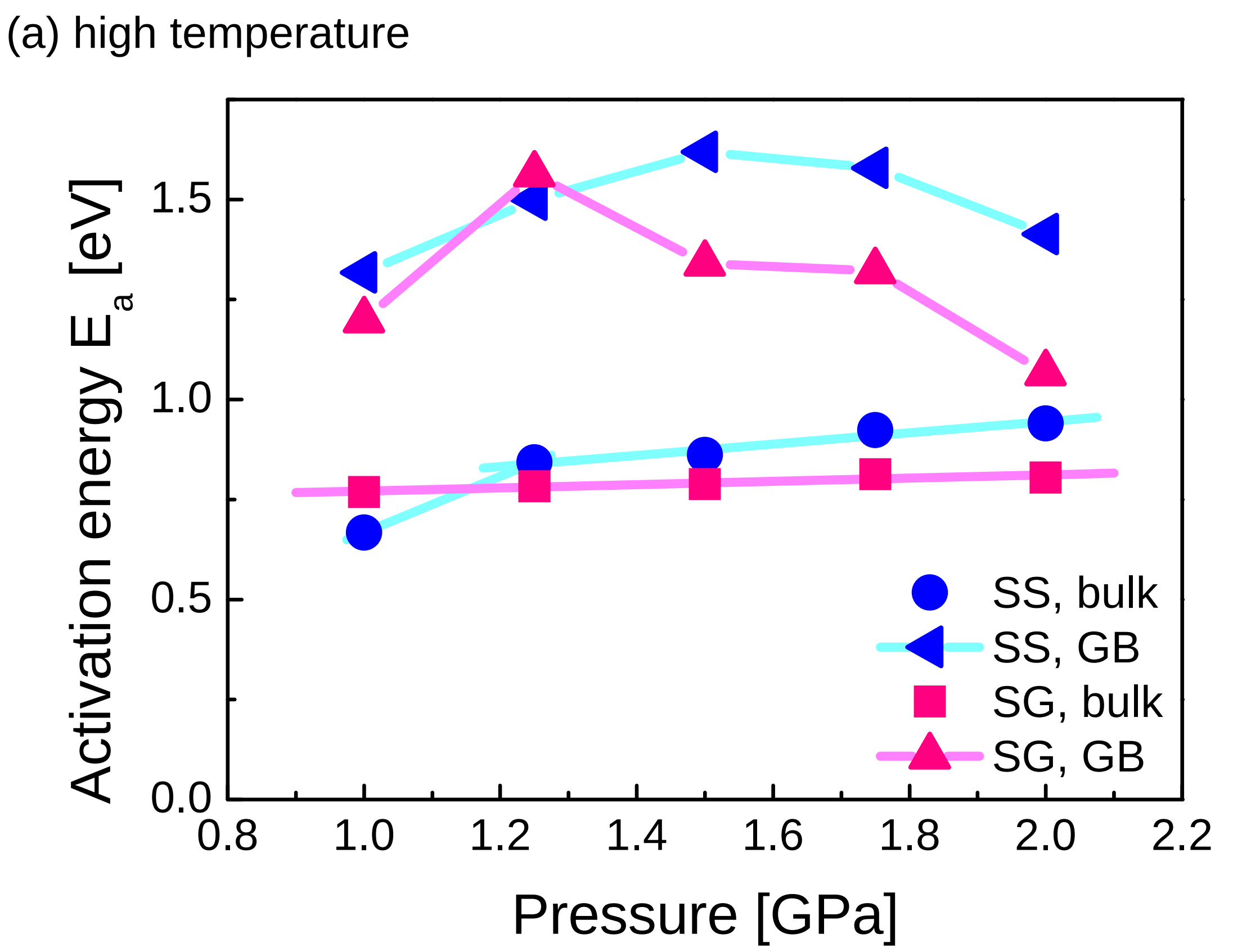

**Figure 4 Activation energy of bulk and grain boundary (GB) conductivity for solid-state and sol-gel synthesized samples at various pressure.**

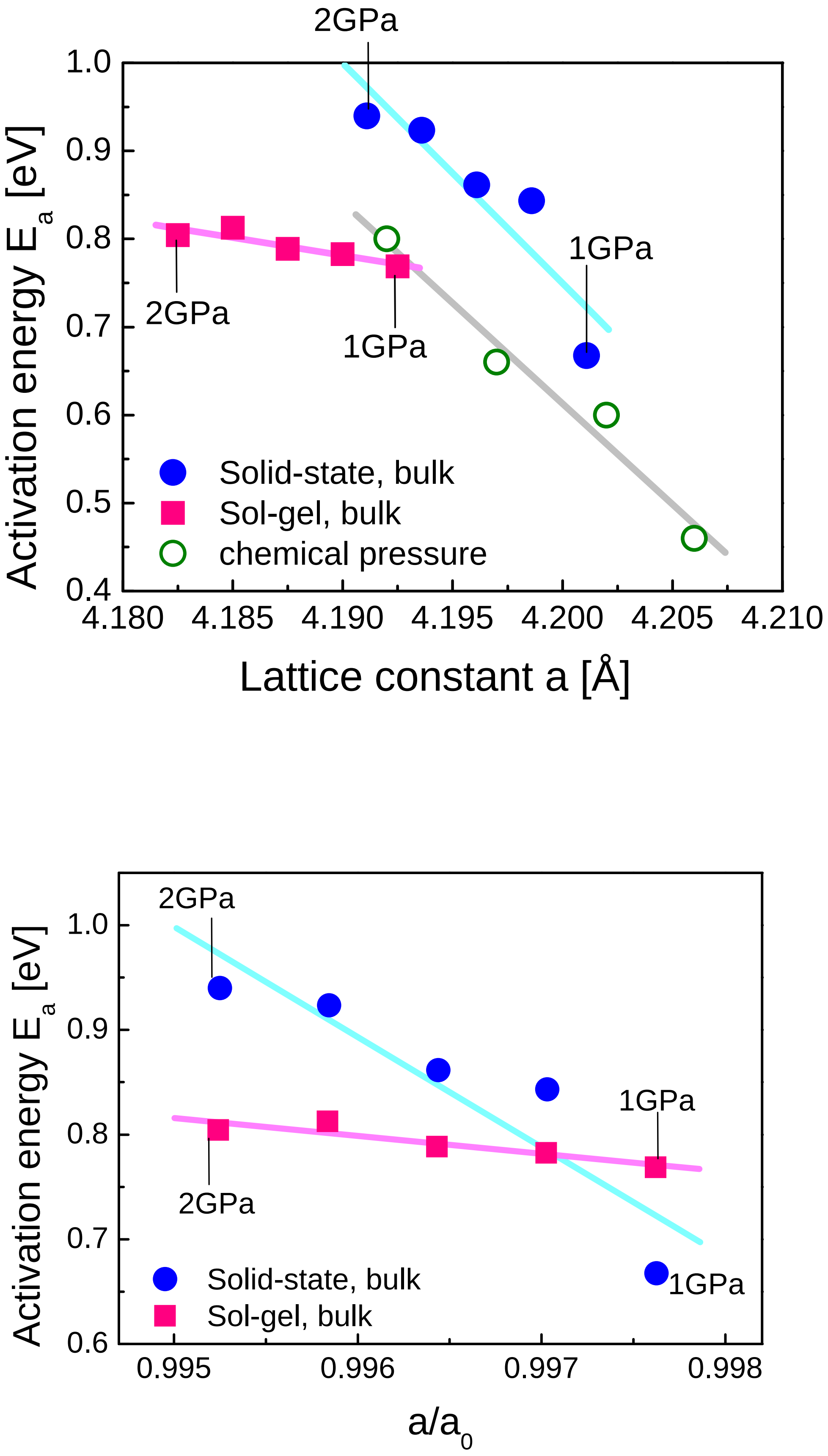

**Figure 5 Comparison of bulk conductivity and activation energy vs. lattice parameter, under mechanical and "chemical" pressure.**